\documentclass[aps,prb,twocolumn,showpacs,preprintnumbers,eqsecnum,amssymb]
{revtex4-1}

\usepackage{amsthm}
\usepackage{amsfonts}
\usepackage{graphicx}

\begin{document}

\title{Delay times in chaotic quantum systems}

\author{A.~M. Mart\'{\i}nez-Arg\"uello}

\author{A.~A. Fern\'andez-Mar\'in}

\author{M. Mart\'{\i}nez-Mares}

\affiliation{Departamento de F\'isica, Universidad Aut\'onoma 
Metropolitana-Iztapalapa, Apartado Postal 55-534, 09340 M\'exico Distrito
Federal, Mexico}

\begin{abstract}
By an inductive reasoning, and based on recent results of the joint moments of 
proper delay times of open chaotic systems for ideal coupling to leads, we 
obtain a general expression for the distribution of the partial delay times for 
an arbitrary number of channels and any symmetry. This distribution was not 
completely known for all symmetry classes. Our theoretical distribution is 
verified by random matrix theory simulations of ballistic chaotic cavities.  
\end{abstract}

\pacs{73.23.-b, 73.23.Ad, 05.45.Mt, 05.60.Gg}

\maketitle

\section{Introduction}
\label{sec:intro}

The delay experienced by a quantum particle due to the interaction with a 
scattering region has been the subject of intense investigation for more than 
thirty years in several areas that include nuclear and condensed matter 
physics.\cite{Wigner,Smith,Lyuboshits,Bauer,Landauer,Price,Iannaccone} The 
interest in this subject has resurged due to the recent appearance of 
theoretical investigations in chaotic 
systems\cite{Berkolaiko,Mezzadri1,Mezzadri2,Kuipers,Marciani,Novaes1,
Novaes2} and atomic physics;\cite{Feist2014,Ivanov2014,Deshmukh2014,Chacon2014} 
the later motivated by experiments of interaction of light with matter during a 
mean time with attosecond precision.\cite{Schultze2010}

The delay time first introduced by Wigner\cite{Wigner} for one channel and 
its multichannel generalization by Smith,\cite{Smith} in the so-called 
Wigner-Smith time delay matrix, is written in terms of the scattering matrix $S$ 
and its derivative with respect to the energy $\varepsilon$. In units of the 
Heisenberg time $\tau_{{}_{\rm{H}}}$, it is given by
\begin{equation}
\label{eq:WignerSmith}
Q_{{}_{\rm{W}}} = -\mathrm{i}\frac{\hbar}{\tau_{{}_{\rm{H}}}}
S^{-1} \frac{\partial S}{\partial\varepsilon}.
\end{equation}
The eigenvalues of $Q_{{}_{\rm{W}}}$ represent the delay time on each channel 
and the Wigner time delay is the average of these proper delay times. In 
the context of mesoscopic systems the electrochemical capacitance of a 
mesoscopic capacitor is described by the Wigner time 
delay.\cite{Lambe,Buttiker1,Buttiker2} Some other transport observables that 
depend on the proper delay times are the thermopower,\cite{vanLangen} the 
derivative of the conductance with respect to the Fermi energy,\cite{BrouwerdG} 
the DC pumped current at zero bias,\cite{BrouwerPumping} among others (see 
for instance Ref.~\onlinecite{BrouwerWRM} and references there in). For 
ballistic systems with chaotic classical dynamics, these physical observables 
fluctuates with respect small variations of external parameters, like an applied 
magnetic field, the Fermi energy or the system 
shape,\cite{FyodorovJMP,Frahm1,BrouwerWRM} the proper delay times are of 
interest in the characterization of their universal statistics. The distribution 
of the proper delay times is given in terms of the joint distribution of their 
reciprocals, known as the Laguerre ensemble,\cite{Frahm1,BrouwerWRM} which 
depends on the particular symmetry present in the problem; an interesting 
feature of this ensemble is the presence of level repulsion, as occur in the 
spectral statistics of several complex quantum-mechanical systems. 

Alternatively, the partial delay times defined as the energy derivative of phase 
shifts are also useful in the characterization of chaotic 
scattering.\cite{FyodorovJMP} Although the partial times are correlated, this 
correlation is of different nature than that between the proper delay times; 
they do not show the level repulsion.\cite{Savin} In the one channel situation 
the proper an partial delay times are identical to the Wigner time delay $\tau$ 
whose distribution is known for all symmetry 
classes:\cite{FyodorovJMP,Gopar,FyodorovPRE55} $\beta=1$ (4) in the presence 
of time reversal and presence (absence) of spin-rotation symmetry, and $\beta=2$ 
in the absence of time reversal symmetry. For the general case of arbitrary 
number of channels, the distribution of a partial time is known for 
$\beta=2$\cite{FyodorovPRL76,Seba} and for $\beta=1$ an expression in terms of 
quadratures was obtained for non-ideal coupling.\cite{FyodorovPRE55} In the 
ideal coupling case only the tails of the distribution are known and the 
corrected prediction does not seem to be right.\cite{Seba} Moreover, the 
$\beta=4$ symmetry is seldom discussed and the distribution of partial times for 
this symmetry class is not given yet. 

In the present paper, we obtain, by an inductive reasoning, a general 
expression for the probability distribution of the partial delay times. This was 
done by extracting the essence that comes from the level repulsion in the joint 
distribution of proper delay times, that transcends to the $k$th moment of a 
proper delay time.\cite{Angeljmp} We test our formula by random matrix theory 
simulations for all symmetry classes and for several number of channels.

In the next section we establish the theoretical framework of the proper and 
partial delay times; we review the known results for the $k$th moment of a 
proper delay time from which we obtain a general expression of the probability 
distribution of the partial times, for all symmetry classes and any number of 
channels. In Sect.~\ref{sect:simulations} we compare our generalized 
distribution with the numerical predictions from random matrix theory. We 
conclude in Sect.~\ref{sect:conclusions}.

\section{Distributions of proper and partial delay times}
\label{sect:times}

\subsection{Scattering approach}

Single-electron scattering by a ballistic cavity attached ideally to two 
leads which support $N_1$ and $N_2$ propagating modes (channels), 
respectively, can be described by a $N\times N$ scattering matrix $S$, where 
$N=N_1+N_2$. When the dynamics of the cavity is classically chaotic, the 
scattering matrix belongs to one of the three circular ensembles of random 
matrix theory (RMT).\cite{Mehta,Dyson} The circular unitary ensemble (CUE) is 
obtained when flux conservation is the only restriction in the problem, such 
that $S^{\dagger}S=\openone_N$, where $\openone_N$ denotes de unit matrix of 
dimension $N$. In the Dyson scheme this case is labeled by $\beta=2$. 
Additionally, in the presence of time reversal invariance (TRI) and integral 
spin or TRI, half-integral spin and rotation symmetry, $S$ is a symmetric 
matrix, $S=S^T$ (the upper script $T$ means transpose). This case is 
denoted by $\beta=1$ and the corresponding ensemble is the circular orthogonal 
ensemble (COE). In the presence of TRI, half-integral spin, and no rotation 
symmetry, $S$ is self-dual and the ensemble of self-dual scattering matrices is 
the circular symplectic ensemble (CSE), labeled by $\beta=4$. In 
the diagonal form, the $S$ matrix can be written as
\begin{equation}
S = U E U^{\dagger}, 
\end{equation}
where $U$ is a $N\times N$ unitary, the matrix of eigenvectors, and $E$ 
is the diagonal matrix of eigenphases, 
\begin{equation}
E_{ij} = \mathrm{e}^{\mathrm{i}\theta_{i}} \, \delta_{ij}, 
\end{equation}
with $\delta_{ij}$ the Kronecker delta. 

\subsection{Proper delay times}

A symmetrized form of the Wigner-Smith time delay matrix can be written in 
dimensionless units as\cite{Frahm1,BrouwerWRM} 
\begin{equation}
Q = -\mathrm{i} \frac{\hbar}{\tau_{{}_{\rm{H}}}}\, S^{-1/2} 
\frac{\partial S}{\partial\varepsilon}
S^{-1/2},
\end{equation}
where $\varepsilon$ is the energy and $\tau_{{}_{\rm{H}}}$ is the Heisenberg 
time ($\tau_{{}_{\rm{H}}}=2\pi\hbar/\Delta$, with $\Delta$ the mean level 
spacing). The matrix $Q$ is Hermitian for $\beta=2$, real symmetric for 
$\beta=1$, and quaternion self-dual for $\beta=4$. Its eigenvalues, $q_i$'s 
($i=1,\,\ldots,\,N$) are the proper delay times measured in units of 
$\tau_{{}_{\rm{H}}}$. The distribution of the $q_i$'s in terms of 
their reciprocals $x_i=1/q_i$ is given by the Laguerre ensemble\cite{Frahm1}
\begin{equation}
\label{eq:Laguerre}
p_{\beta}(\{x_i\}) = C_N^{(\beta)} 
\prod_{a<b}^N \left| x_b - x_a \right|^{\beta} 
\prod_{c=1}^N x_c^{\beta N/2}\, \mathrm{e}^{-\beta x_c/2}  ,
\end{equation}
where $C_N^{(\beta)}$ is a normalization constant. It is worth mentioning that 
the level repulsion that appears between the proper delay times is inherited of 
the Hamiltonian eigenvalues. There, its normalization constant is well 
known;\cite{Guhr} however, the constant $C_N^{(\beta)}$ has not been given yet, 
although the Laguerre distribution has been widely used. Through the analysis 
given in Ref.~\onlinecite{Angeljmp} we find a general expression given by 
\begin{equation}
C^{(\beta)}_N = 
\frac{\left[ \left(\frac{\beta}{2}\right)^{\beta(N-1/2)+1} 
\left(\frac{\beta}{2}\right)! \right]^N}
{\left(\frac{\beta N}{2}\right)!} 
\prod_{n=0}^{2N-1} \frac{1}{\left(\frac{\beta n}{2}\right)!}\, .
\end{equation}

Let us note that the level repulsion of the proper delay times in 
(\ref{eq:Laguerre}) transcends to the $k$th moment of them; recent results of 
that $k$th moment, valid for any symmetry and an arbitrary number of 
channels, shows the underlying part that comes from this 
repulsion, namely\cite{Angeljmp}
\begin{equation}
\label{eq:qNk}
\left\langle q_i^k \right\rangle^{(\beta)} = 
\left(\frac{\beta}{2}\right)^k 
\frac{\left(\frac{\beta N}{2}-k\right)!}{\left(\frac{\beta N}{2}\right)!}\,
K_N^{(\beta)}(k,0,\ldots,0),
\end{equation}
for $k<1+\beta N/2$. The factor $K_N^{(\beta)}(k,0,\ldots,0)$ is the 
inheritance part of the level repulsion according to the above statement. For 
$k=1$, 2, and 3 it becomes independent of $\beta$,\cite{Angeljmp} 
\begin{equation}
\label{eq:Kk123}
K_N^{(\beta)}(k,0,\ldots,0) = \frac{k!N^{k-1}N!}{(N+k-1)!}.
\end{equation}

\subsection{Partial delay times}

Partial delay times, defined as the energy derivative of the diagonal form of 
the scattering matrix as in Eq.~(\ref{eq:WignerSmith}), are given, in 
dimensionless units, by\cite{FyodorovJMP,Seba} 
\begin{equation}
\hat{\tau} = -\mathrm{i} \frac{\hbar}{\tau_{{}_{\rm{H}}}} E^{-1}
\frac{\partial E}{\partial \varepsilon}. 
\end{equation}
It is an $N\times N$ diagonal matrix with elements 
\begin{equation}
\tau_s = \frac{\hbar}{\tau_{{}_{\rm{H}}}} 
\frac{\partial \theta_s}{\partial\varepsilon}.
\end{equation}

Since the partial delay times dot not show level repulsion, once the inherent 
part of the level repulsion has been identified, it is easy to arrive at the 
expression of the $k$th moment of the partial time, that is\cite{Angeljmp}
\begin{equation}
\label{eq:qNkpartial}
\left\langle \tau_s^k \right\rangle^{(\beta)} = 
\left(\frac{\beta}{2}\right)^k 
\frac{\left(\frac{\beta N}{2}-k\right)!}{\left(\frac{\beta N}{2}\right)!}, 
\end{equation}
This expression is in agreement with the results that can be obtained directly  
from the distribution\cite{Gopar} for the $N=1$ case and any $\beta$. Also, 
Eq.~(\ref{eq:qNkpartial}) is in agreement with the known result for 
$\beta=2$ and arbitrary $N$.\cite{FyodorovJMP,FyodorovPRL76} Following this 
inductive line of thought, we arrive at the distribution of the partial delay 
times for all symmetry classes and any number of channels, namely
\begin{equation}
\label{eq:Ps}
P_{\beta}(\tau_s) = \frac{2/\beta}{\left(\frac{\beta N}{2}\right)!}\, 
\left( \frac{\beta}{2\tau_s} \right)^{2+\beta N/2} 
\mathrm{e}^{-\beta/2\tau_s}.
\end{equation}
This is our main result in this paper. Our expression encompasses the existing 
results in the 
literature.\cite{Gopar,FyodorovPRE55,FyodorovJMP,FyodorovPRL76,Seba} 

In what follows we verify our finding with random matrix theory simulations.


\section{Numerical calculations}
\label{sect:simulations}

The Hamiltonian approach, also known as the Heidelberg approach, is the 
best suited for the calculation of the energy derivative of the scattering 
matrix since it is written explicitly in terms of the energy, 
namely~\cite{Guhr,Verbaarschot,BeenakkerRMP}
\begin{equation}
S(\varepsilon) = \openone_N - 2{\rm i}\pi W^{\dagger} 
\frac{1}{\varepsilon \openone_{M}-H+{\rm i}\pi WW^{\dagger}} W, 
\end{equation}
where $H$ is an $M$-dimensional Hamiltonian matrix that describes the chaotic 
dynamics of the system, with $M$ resonant single-particle states, and $W$ is a 
$M\times N$ matrix, independent of the energy, which couples these resonant 
states with the $N$ propagating modes in the leads; $\openone_n$ stands for the 
unit matrix of dimension $n$. For ideal coupling of uncorrelated equivalent 
channels, $W_{\mu n}=\sqrt{M\Delta}/\pi$ ($\mu=1,\,\ldots,\,M$ and 
$n=1,\,\ldots,\,N$) for the matrix elements of $W$.\cite{BeenakkerRMP}

\begin{figure}
\includegraphics[width=1.0\columnwidth]{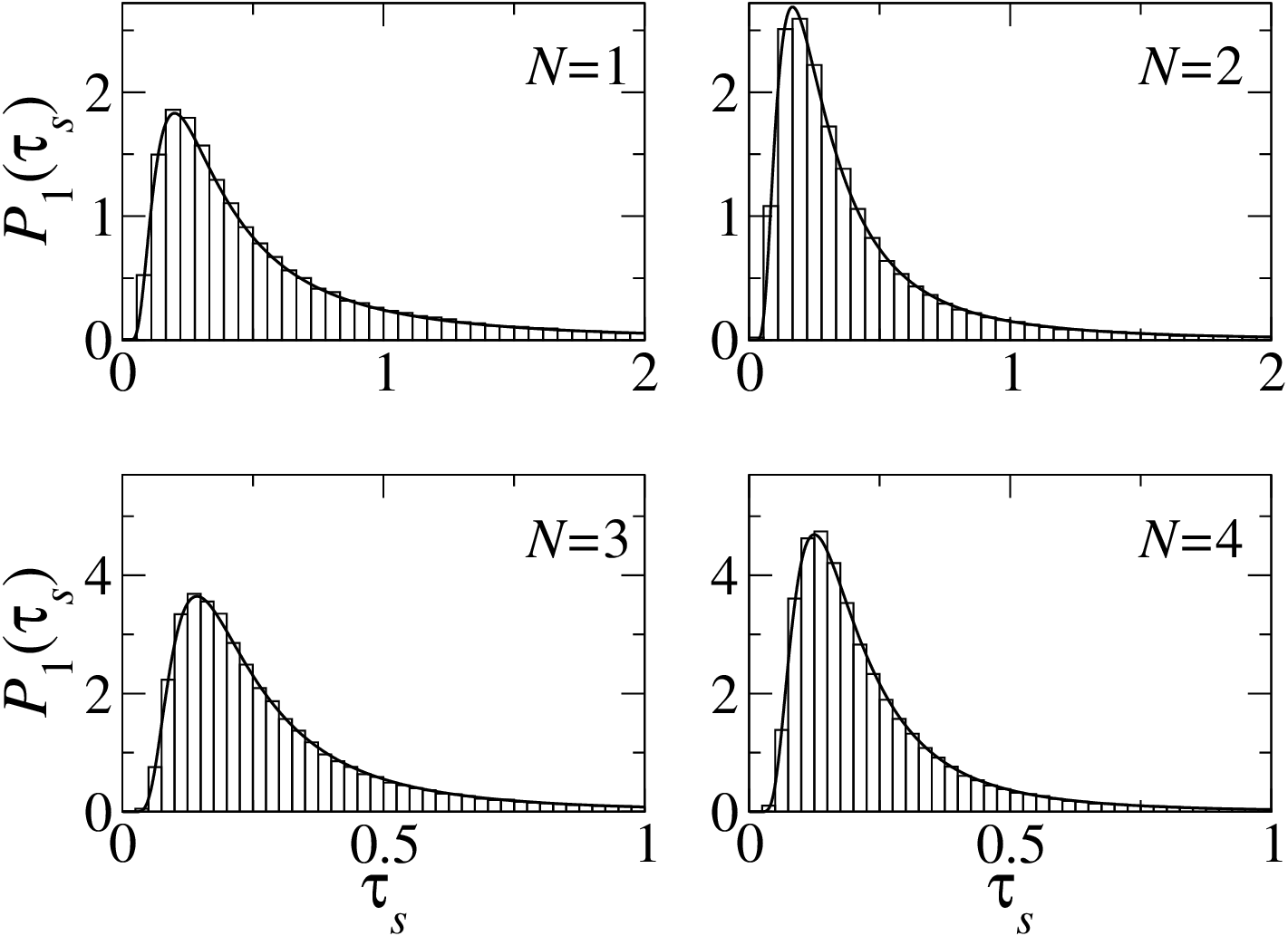}
\caption{Comparison between the numerical simulations (histograms) and theory 
(continuous lines), Eq.~(\ref{eq:Ps}), for the distribution of $\tau_s$ (we 
take $s=1$) in the $\beta=1$ case.}
\label{fig:beta1}
\end{figure}

For chaotic systems, $H$ is a random matrix chosen from one of the Gaussian 
ensembles: orthogonal ($\beta=1$), unitary ($\beta=2$) or symplectic 
($\beta=4$). The matrix elements of $H$ are uncorrelated random variables with 
a Gaussian probability distribution with zero mean and variance 
$\lambda^2/\beta M$; the later determines the mean level spacing at the 
center of the band, $\Delta=\pi\lambda/M$.\cite{Guhr} An ensemble of 
Hamiltonian matrices leads to an ensemble of $S$-matrices, which represents the 
several realizations of systems for which the statistical analysis is 
performed. To implement the simulations we followed the same method as in 
Ref.~\onlinecite{Mucciolo} for $\beta=1$ and 2, while for $\beta=4$ the 
subroutine given in Ref.~\onlinecite{Cappellini} was used to generate the 
random Hamiltonian.

For each realization we diagonalize the matrix $S$ to determine its eigenvalues. 
We are interested in one of the eigenvalues only, 
$E_s(\varepsilon)=\exp[\mathrm{i}\theta_s(\varepsilon)]$ let say, but 
evaluated at three energies in order to calculate the energy derivative. That 
is,  
\begin{equation}
\tau_s = -\frac{\mathrm{i}}{2\pi\epsilon} \,
\frac{E_s(\epsilon/2)-E_s(-\epsilon/2)}{E_s(0)}, 
\label{eq:taus-numeric}
\end{equation}
where $\epsilon=\varepsilon/\Delta$.

\begin{figure}
\includegraphics[width=1.0\columnwidth]{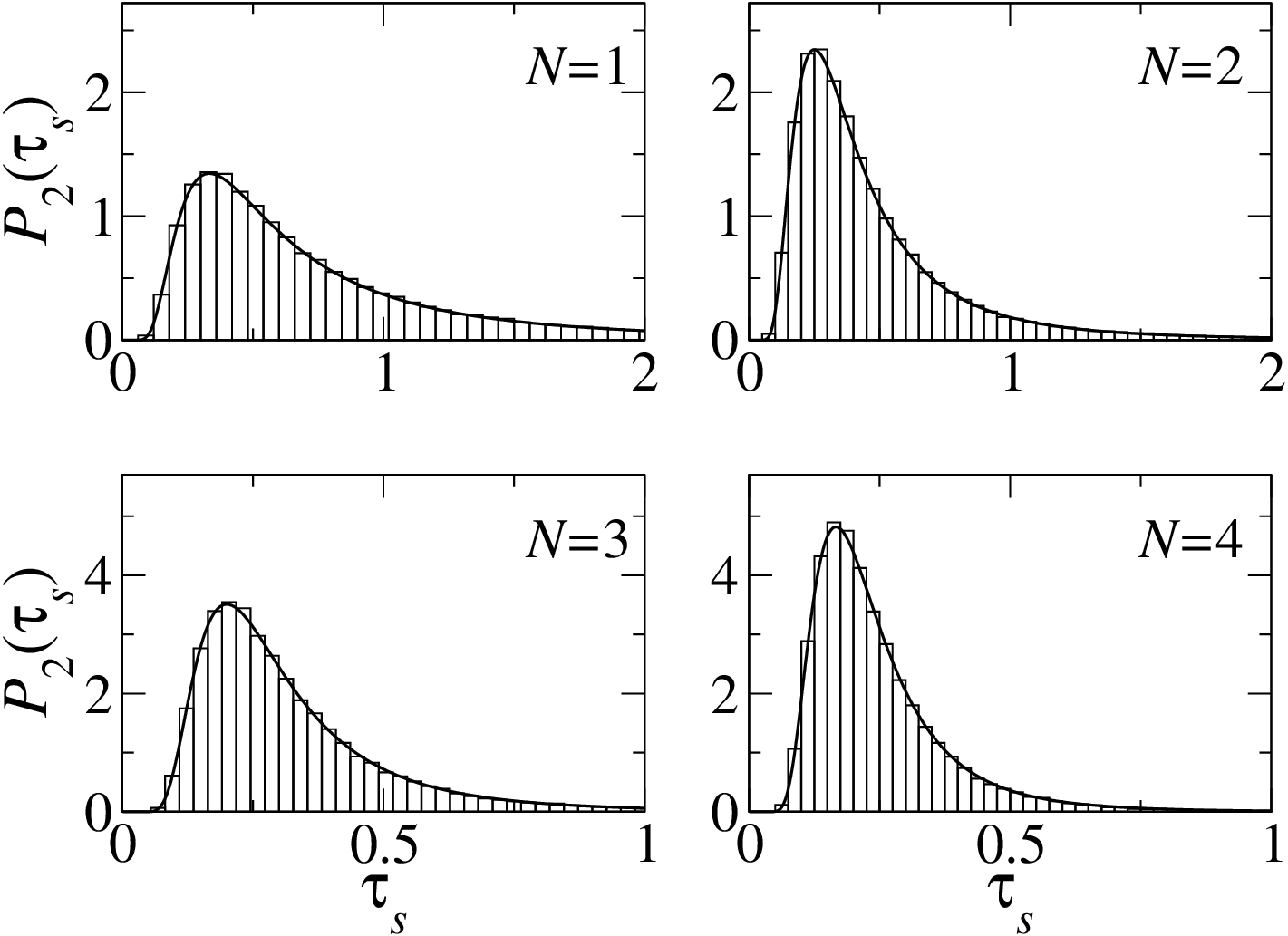}
\caption{The same as in Fig.~\ref{fig:beta1} but for $\beta=2$.}
\label{fig:beta2}
\end{figure}

\begin{figure}
\includegraphics[width=1.0\columnwidth]{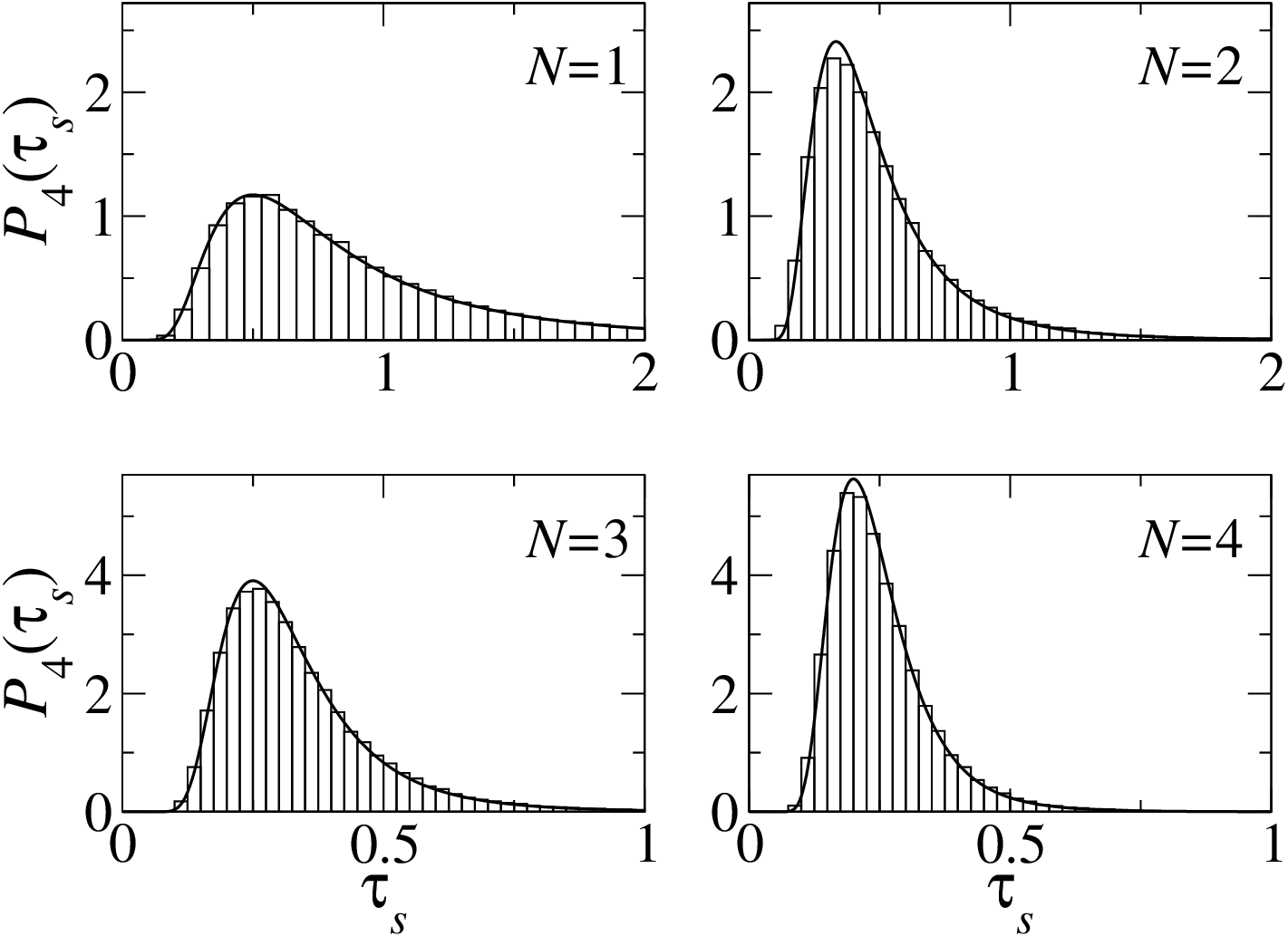}
\caption{The same as in Fig.~\ref{fig:beta1} but for $\beta=4$.}
\label{fig:beta4}
\end{figure}

In Fig.~\ref{fig:beta1} we compare our theoretical distribution for the partial 
delay times, Eq.~(\ref{eq:Ps}), for $\beta=1$, with the numerical results 
obtained from the random matrix simulations with $10^5$ realizations of 
$\tau_s$, calculated as in Eq.~(\ref{eq:taus-numeric}) for $M=100$ and 
$\epsilon=0.001$. We observe an excellent agreement for the several cases of $N$ 
presented. This result is an important one since it has not been verified 
before. Figure~\ref{fig:beta2} shows the corresponding results for $\beta=2$, 
which is in agreement those of Ref.~\onlinecite{Seba}. For $\beta=4$ the 
comparison is shown in Fig.~\ref{fig:beta4} where also we observe an excellent 
agreement. Let us note that this is the first time that the distribution of the 
partial times is given and verified for the $\beta=4$ and any number of 
channels. 

\section{Conclusions}
\label{sect:conclusions}

Based on known results of the joint moments of proper delay times, we 
obtained the distribution of the partial delay times for an arbitrary number of 
channels and any symmetry. This was done following and inductive method by 
extracting the underlying part coming from the level repulsion inherited to 
the $k$th moment of a proper delay time. Our distribution not only reproduces 
theoretical results previously considered in the literature for unitary 
symmetry, but also extends it to the orthogonal and symplectic symmetries, 
for an arbitrary number of channels. Besides, we were able to provide the 
normalization constant for the joint distribution of proper delay times. We 
tested our theoretical distribution by random matrix theory simulations of 
ballistic chaotic cavities with ideal coupling. 

\acknowledgments

A.~M. Mart\'inez-Arg\"uello thanks CONACyT, Mexico, for financial support. 
A.~A. Fern\'andez-Mar\'in also thanks financial support from PRODEP under the 
project No. 12312512. M. Mart\'inez-Mares is grateful with the Sistema Nacional 
de Investigadores, Mexico.



\end{document}